Silver staining of proteins in polyacrylamide gels

Mireille Chevallet, Sylvie Luche and Thierry Rabilloud*
CEA-Grenoble, DRDC/ICH; INSERM U548
17 rue des martyrs, 38054 Grenoble Cedex 9

* corresponding author
email: thierry.rabilloud@cea.fr
Phone +33 438 783 212, Fax : +33 438 789 803

**Abstract**

Silver staining is used to detect proteins after electrophoretic separation on polyacrylamide gels. It combines excellent sensitivity (in the low nanogram range) whilst using very simple and cheap equipment and chemicals.  It is compatible with downstream processing such as mass spectrometry analysis after protein digestion. The sequential phases of silver staining are protein fixation, then sensitization, then silver impregnation and finally image development. Several variants of silver staining are described here, which can be completed in a time range from 2 hours to one day after the end of the electrophoretic separation. Once completed, the stain is stable for several weeks

**Introduction**

Among the various protein detection methods following electrophoresis of polyacrylamide gels, silver staining has gained wide popularity because of its sensitivity (in the very low ng range),  because it can be achieved with simple and cheap laboratory reagents, and because it does not require complicated and expensive hardware for the readout. Furthermore, silver staining, or at least some of its variants, is also compatible with downstream processing such as mass spectrometry

The rationale of silver staining is quite simple. Proteins bind silver ions, which can be reduced under appropriate conditions to build up a visible image made of finely divided silver metal. However, silver staining can be a tricky business, as many artefacts and pitfalls exist. Unlike staining with organic dyes, silver staining goes against general thermodynamics. Proteins bind silver ions and this binding decreases the reactivity of the ions. This general phenomenon explains why "hollow" or "doughnut" bands or spots are so commonly encountered. However, silver ion reduction is extremely self-catalytic, so that any trick that promotes silver reduction at the site of proteins will favor a sensitive and positive staining of proteins. The description and discussion of those tricks, which are part of the sensitization process, are beyond the scope of this paper and can be found elsewhere [1].

All silver staining protocols, starting form the pioneer work [2] are made of the same basic steps, which are: i) fixation to get rid of interfering compounds; ii) sensitization and rinses to increase the sensitivity and contrast of the staining; iii) silver impregnation with either a silver nitrate solution or a silver-ammonia complex solution; iv) rinses and development to build up the silver metal image; and v) stop and rinse to end development prior to excessive background formation and to remove excess silver ion and other chemicals prior to further processing.

Because of the variations that can be introduced at each of these steps, several dozens of protocols are described in the literature, which can be somewhat confusing. It is therefore the purpose of this article to exemplify selected protocols which face the most commonly encountered constraints.

In the field of silver staining, two families of methods are co-existing, depending of the reagent used for silver impregnation. This can be either simple silver nitrate, or a silver-ammonia complex.
The silver nitrate stains are the simplest ones. The compatibility with mass spectrometry is moderate, but can be improved with some modifications in the fast track (listed below), at the expense of the quality of the stain itself. More background is obtained at high room temperatures (>30°C). Basic proteins are less efficiently stained than acidic ones with this type of stain. This type of stain is also compatible with various electrophoretic systems (including for example the Tricine [3] and Bicine [4] systems and works for most commercial gels. This stain is a good choice when numerous gels must be stained and when staining consistency is important.
This silver-ammonia stains are less straightforward than silver nitrate stains, but they offer more flexibility in the control of staining. They do not work properly below 19-20°C, except when the water used for making the solutions from steps is warmed at 20-25°C or above prior to use. This type of stain stains basic proteins more efficiently than acidic ones. Only the classical glycine electrophoresis system and the taurine system [5] can be used with this type of stain, which require home-made gels [6]. Gels must be polymerized with a complex initiator system. To 1 ml of gel mix, 0.7 µl of TEMED, 5µl of 10% sodium thiosulfate pentahydrate, and 7µl of 10% ammonium thiosulfate are sequentially added. The gel is then cast and polymerized for at least one hour. Gels made in this way can be kept in a cold room for up to a week.

When selecting the most appropriate protocol, the experimenter must answer a few questions:

**1) Do I want to use only precast gels or am I ready to cast my own polyacrylamide gel ?**
Some very efficient silver staining protocols (e.g. those using silver-ammonia complex) require special gels which can be easily made in the laboratory, but which have a limited shelf life, preventing them from being commercialized.

**2) Will I stain 1D or 2D gels ?**
2D gels are often contaminated with carrier ampholytes, which are difficult to remove and give staining artefacts. Consequently, 2D gels require more rigorous fixation to remove thoroughly the ampholytes

**3) Do I wish to go further with my spots (e.g. with mass spectrometry) or is silver staining the final step ?**
Fixation with aldehydes (formaldehyde, glutaraldehyde) dramatically improves fixation, sensitivity and uniformity of staining. It however precludes any further use of such stained spots.

**4) Is downstream processing more important than silver staining itself ?**
Roughly speaking, there is a balance between the quality of staining itself and the yields obtained in downstream processing. Steps favoring a better image (e.g. increased fixation) decrease the peptide yield in mass spectrometry.

**5) Which parameter do I favor between speed and regularity of staining ?**
Generally speaking, fast protocols use short steps (one minute or less), which are difficult to make very reproducible. This may be a problem when staining consistency is a critical parameter (e.g. in large series of 2D gels)

**6) What is the quality of temperature control in the laboratory ?**
As silver staining is a delicate process, its is quite temperature dependent. Consequently, some protocols work poorly when it is too cold (below 20°C) while other protocols work poorly when it is too hot ( above 30°C).

Depending on the answers to these questions, one protocol will be more optimal than the others.
Typical situations and corresponding choices are the following:

-Fast, sensitive visualization of proteins, on only a few gels per series,  no special needs afterwards: select fast silver staining (protocol A)
-comparative studies of large series of gels over an extended time period, maximal consistency and good linearity of stain required: select long silver nitrate staining (protocol B)
-sensitive detection required, without any quantitative analysis, but maximal sequence coverage required in subsequent analysis with mass spectrometry: select ultrafast silver nitrate staining (variation of protocol A)
-combination of sensitivity, linearity, and good compatibility with mass spectrometry: select aldehyde-free silver ammonia staining (protocol C)
-maximal sensitivity required, no need for any subsequent analysis: select formadehyde-silver-ammonia staining (protocol D)

Note that not all the reagents and materials listed below are necessary for a given silver staining protocol. Check what is needed by reading the procedure first.

**General practice**

Batches of gels (up to four gels per box) can be stained. For a batch of three to five medium-sized gels (e.g. 160 x 200 x 1.5 mm), 1 l of the required solution is used, which corresponds to a solution/gel volume ratio of  5 or more. 500 ml of solution is used for one or two gels of this size. the volumes can be adjusted according to the gel size, provided that a solution/ gel volume ratio is at least 5 and that the gel(s) float(s) freely in the solution. Batch processing can be used for every step longer than 5 min, except for image development, where one gel per box is required (the steps where batch processing cannot be used are stated). For steps shorter than 5 min, the gels should be dipped individually in the corresponding solution.

For changing solutions, the best way is to use a rigid plastic sheet. The sheet is pressed on the pile of gels with the aid of a gloved hand. Inclining the entire setup allows the emptying of the box while keeping the gels in it. The next solution is poured with the plastic sheet in place, which prevents the incoming solution flow from breaking the gels. The plastic sheet is removed after the solution change and kept in a separate box filled with water until the next solution change. This water is changed after each complete round of silver staining.

When gels must be handled individually, they are manipulated with gloved hands. The use of powder--free, nitrile gloves is strongly recommended, as powdered latex gloves are often the cause of pressure marks. Except for development or short steps, where occasional hand agitation of the staining vessel is convenient, constant agitation on a rocking table is required for all the steps.

Four different silver staining protocols are described in this paper, for improving the coverage of the various experimental needs encountered.

**Materials**

**Reagents:**

Acetic acid (glacial)

Ethanol: CRITICAL: 95% ethanol can be used instead of absolute ethanol without any volume correction. it is strongly recommended NOT to use denatured alcohol.

Water: CRITICAL: silver staining is very sensitive to trace impurities in water. Water with a resistivity greater than 15MOhm/cm must be used

Silver nitrate. CRITICAL: This is usually used as a stock solution (typically 20% w/v), which must be stored in a dark place (a fridge is an ideal cold, dark place. Otherwise, a titrated silver nitrate solution (1M, available from e.g. Riedel de Haën) can be purchased and used

Ammonium hydroxide: Here again, the use of a titrated ammonium hydroxide solution (5M, from Aldrich) is strongly recommended. This stock solution is stored in the fridge and can be stable for months.

Sodium thiosulfate. This chemical is used as the pentahydrate salt, and can be purchased from various suppliers. A good pro analysis grade should be used. CRITICAL: It must be remembered that thiosulfate is not a very stable chemical. Upon aging of solutions or even the stock powder, a yellowish background has a tendency to increase on the gels. For convenience, a 10% stock solution can be prepared. It is stored at room temperature and kept for no longer than a week.

Naphtalene disulfonate: this chemical exists as several isomers, either in the free acid form or in the disodium salt form. The use of 2,7 naphtalene disulfonic acid disodium salt (Acros) is recommended, but 1,5 naphtalene disulfonic acid (Aldrich) can also be used;

Formaldehyde: a 37 or 40% stock solution is used (Formalin). CAUTION: fconcentrated formaldehyde is toxic, and should be handled under a hood. CRITICAL: Formaldehyde polymerizes over time and in the cold, resulting in the decrease in the concentration of active formaldehyde. The formalin solution must NOT be stored at low temperatures, and solution with a heavy deposit of polymer must NOT be used.

Tris
Potassium carbonate
Potassium tetrathionate
Potassium acetate
Ethanolamine
Potassium ferricyanide
Ammonium hydorgenocarbonate

**Equipment**

Rocking table. This should NOT be an orbital shaker, but rather a rocking table with a ping-pong move. Its speed should be adjustable, and it should be able to operate at 30-60 strokes/minute

Containers: Glass or plastic can be used. Glass is easier to clean but heavier and more fragile. Plastic has the opposite characteristics. Among plastics, polyethylene food boxes are recommended. CRITICAL: Plastic boxes should be thoroughly cleaned with acetone and then alcohol prior to first use to remove traces of plasticizers and unmolding agents. The bottom area of the container should be at least 20% greater than the area of the gels to be stained.

Rigid plastic sheet: this is a very useful accessory to change the solutions without touching the gels. We use the polycarbonate sheets available from electrophoresis setups suppliers (e.g. BioRad, GE healthcare) for separating gel units in multi-gel casting chambers.

REAGENT SETUP

**Tetrathionate sensitizing solution**:
For a liter of solution, dissolve 50g of potassium acetate and 3 g of potassium tetrathionate in 500ml of water. Add water to 700ml, then 300ml of ethanol.
CAUTION: this solution must be prepared the day of use

**Silver-ammonia solution.**
The silver-ammonia solution is prepared as follows: for ca. 500 ml of staining solution, 475 ml of water is placed in a flask with strong magnetic stirring. First, 7 ml of 1 N sodium hydroxide is added, followed first by 7.5 ml of 5 N ammonium hydroxide (Aldrich) and then 12 ml of 1 N silver nitrate. A transient brown precipitate forms during silver nitrate addition. It should disappear a few seconds after the end of silver addition. Persistence of a brown precipitate or color indicates exhaustion of the stock ammonium hydroxide solution. Attempts to correct the problem by adding more ammonium hydroxide are useless as sensitivity drops.
CRITICAL  The ammonia--silver ratio is a critical parameter for good sensitivity [7]. The above proportions give a ratio of 3.1, which is one of the lowest practical ratios. For special purposes, e.g. when a stain with a sensitivity intermediate between Coomassie blue and classical silver staining is needed, the ammonia concentration can be increased up to 2.5 fold above the given concentrations, resulting in "derating" the silver staining.

CAUTION:  Flasks used for preparation of silver--ammonia complexes  and silver--ammonia solutions must not be left to dry out, as explosive silver azide may form. Flasks must be rinsed at once with distilled water, while used silver solutions should be put in  a dedicated waste vessel containing either sodium chloride or  a reducer (e.g. ascorbic acid) to precipitate silver.

**Developer for silver nitrate staining** (basic developer): 3% potassium carbonate plus 250 µl formalin and 125 µl 10% sodium thiosulfate per liter. CAUTION: developer must be prepared the day of use, and formaldehyde should be added to the developer at most one hour before use.

**Developer for silver ammonia staining** (acidic developer): 350µM citric acid containing 1 ml formalin per liter. Image development but also background development are much faster than with silver nitrate protocols. Thiosulfate, which is a powerful background reducer, cannot be used in the developer and must be included in the gel itself.
CAUTION: developer must be prepared the day of use, and formaldehyde should be added to the developer at most one hour before use

**Stop solution for silver nitrate staining** (Tris stop solution):
4% (w/v) Tris and 2% (v/v) acetic acid. CAUTION: must be prepared the day of use

**Stop solution for silver nitrate staining** (EA stop solution):
0.5% (v/v) ethanolamine and 2% (v/v) acetic acid. CAUTION: must be prepared the day of use. Do NOT mix the pure chemicals and dilute with water (risk of strong exothermic reaction with spillovers) . Dilute first ethanolamine in water, then add acetic acid. Some white fumes of ethanolamine acetate may occur at this stage, but this is without consequences

**Destaining solution for mass spectrometry**
Prepare a 30mM potassium ferricyanide solution in water. Additionally, prepare a 100mM sodium thiosulfate solution. CAUTION: both solutions must be prepared the day of use.
Just before use, mix equal volumes of the potassium ferricyandie and thiosulfate solutions. CAUTION: This resulting yellowish solution is stable and active for less than 30 minutes, and must be used immediately

Procedure

There are four possible methods to fix and stain the gel: short silver nitrate staining (option A); long silver nitrate staining (option B); aldehyde free silver ammonia staining (option C); and silver ammonia staining with formaldehyde fixation (option D).The aldehyde free silver free ammonia staining protocol affords superior compatibility with mass spectrometry [8]. However, it is quite prone to the "hollow spot" phenomenon, especially for acidic glycoproteins (Figure 1). The silver ammonia staining with formaldehyde fixation protocol affords the best sensitivity, uniformity of staining (acidic and basic proteins) and the least chromatism of all silver staining protocols. Unfortunately, it is incompatible with any downstream process.

**A) Short silver nitrate staining** [9,10]  (Timing 2-5 hours).

**i**) After electrophoresis, fix the gels in 30% ethanol, 10% acetic acid for at least 30 minutes.

Pause point: fixation can be carried out for up to 24 hours, with at least one solution change which can take place when desired. However , the first bath of fixation must last at least 30 minutes

Critical step: various fixation schemes can be used, with an influence on the final result. Ultrafast fixation (one 30 minutes bath) improves sequence coverage in subsequent mass spectrometry, but at the expense of a strong chromatism (bands or spots can be yellow, orange, grey or brown) precluding any image analysis. Longer fixation (typically 3 x 30 minutes) gives a much better staining, but a lower sequence coverage. However, this intermediate fixation scheme does not remove adequately carrier ampholytes, which produce a strong background in the low molecular weight region of the gel (Figure 1). Fixation for at least 18 hours is required to remove carrier ampholytes.

**ii**) Rinse the gels twice in 20% ethanol, for 10 minutes for each wash, and then twice in water, for 10 minutes for each wash.

**iii**) Sensitize the gels by soaking (one at a time only) for one minute in 0.8 mM sodium thiosulfate (0.02% if the pentahydrate salt is used)

**iv**) Rinse the gels twice for 1 minute for each wash in water

Critical Step: The optimal setup for sensitization is the following:  prepare four staining boxes containing respectively the sensitizing thiosulfate solution, water (two boxes), and the silver nitrate solution. Put the vessel containing the rinsed gels on one side of this series of boxes. Take one gel out of the vessel and dip it in the sensitizing and rinsing solutions ( 1 min in each solution). Then transfer to silver nitrate. Repeat this process for all the gels of the batch. A new gel can be sensitized while the former one is in the first rinse solution, provided that the 1-minute time is kept (use a bench chronometer).

**v)** Impregnate with 12mM silver nitrate. gels can turn yellow at this stage, but this does not affect the final quality.

pause point: silver nitrate impregnation can last from 20 minutes to 2 hours without any real change in the quality of the results.

**vi)** Only one gel per tray must be used at this stage. The best setup is to arrange on the bench the box containing the gel(s) soaking in silver nitrate, one box half-filled with water, as many boxes containing developer as gels present in the same box of silver solution, plus one box containing the stop solution. The stop solution, made the day of use, contains 40g of tris and 20ml of acetic acid per liter. With gloved hands, rinsed with deionized water, pull out one gel from the silver solution. Dip it for 10 seconds in the water bath, the pull it out from water and transfer it in the developer solution. A brown or grey precipitate normally develops within a few seconds when the gel is dipped in the developer. This precipitate must be redissolved by shaking of the developer-gel- containing box, other wise a particulate surface background will deposit on the gel surface. When this precipitate has redissolved, repeat the procedure for the next gel of the batch. With this protocol, the most intense bands or spots take a few minutes to appear (up to 5 minutes depending on the temperature). Development can be let to proceed up to background development, but nothing happens after 45 minutes of development. As a special trick for one application, when silver staining is used just for track visualization in ID gels prior to systematic band excision, the formaldehyde in the developer can be replaced by 0.5 mM carbohydrazide (final concentration in developer). The image develops and the background turns brown in a minute or two. The reaction is stopped with the usual Tris acetate solution. This protocol offers a terrible staining but the best performance in the subsequent mass spectrometry [11]

**vii)** When the adequate degree of staining has been achieved, transfer the gel to the Tris stop solution for at least 30 minutes. Up to four gels can be piled in one box of stop solution.
pause point: time in stop solution can be extended to up to 2 hours

**viii)** wash gels in water (at least, twice, for 30 minutes for each wash). Stained gels can be stored in water for several days, but when mass spectrometry analysis is forecast, much better results are obtained if the staining, washing, spot excision and destaining are performed on the same day [11].

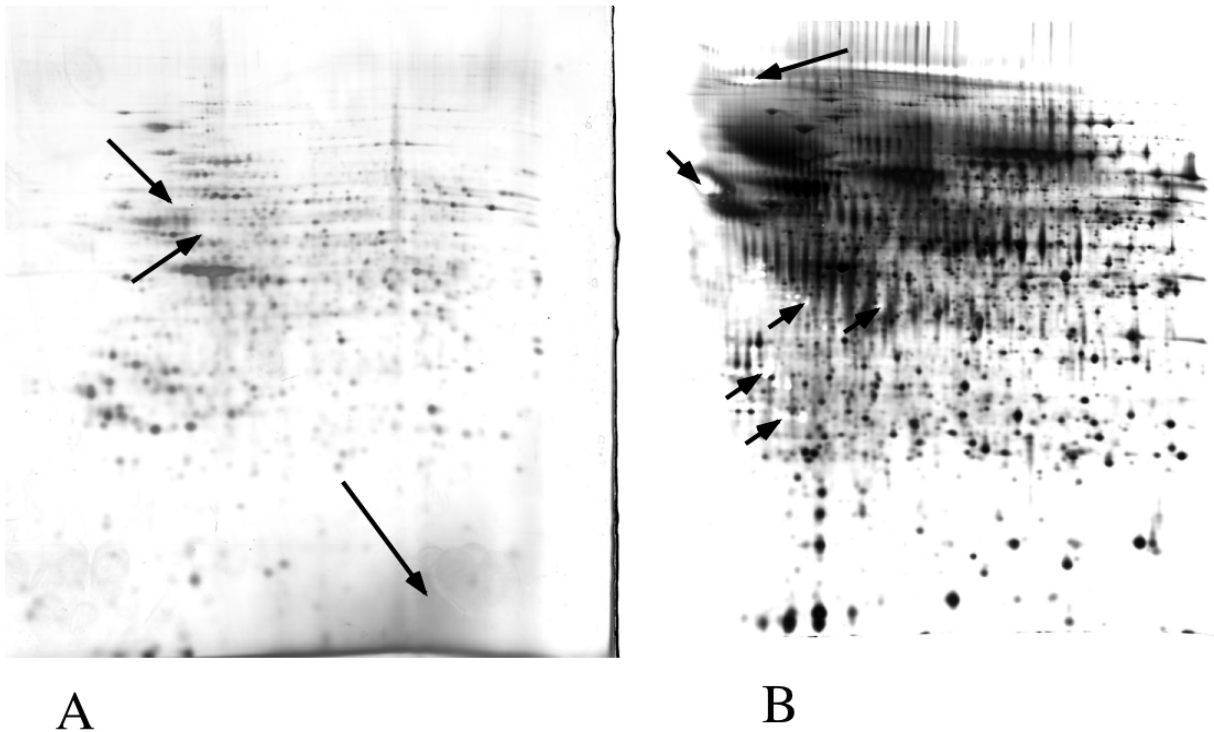

Figure 1 : Examples of staining artefacts
A) HeLa total cell proteins (80µg) were separated by 2D electrophoresis (pH 4-8 gradient in the first dimension10% acrylamide for the second dimension.
Detection by ultrafast silver staining using silver nitrate. Arrows point at weakly stained spots (i.e. very yellow and thus very pale upon scanning in black and white), and at the low molecular weightgrey zone corresponding to carrier ampholytes

B) Monocyte-secreted proteins (150µg) were separated by 2D electrophoresis (pH 4-8 gradient in the first dimension10% acrylamide for the second dimension.
Detection by mass spectrometry-compatible silver-ammonia staining. Arrows point at some of the hollow spots (i.e. completely transparent zones on the gels), typical of this stain

**B) Long silver nitrate staining [12] (Timing overnight + 5 hours)**

**i)** After electrophoresis, fix the gels in 30% ethanol, 10% acetic acid, for 1 hour . then renew the fixation bath and leave to fix overnight

**ii**) Sensitize for 45 minutes in tetrathionate sensitizing solution

**iii**) Rinse twice in 20% ethanol, for 10 minutes for each wash.

**iv**) Rinse four times in water, for 10 minutes for each wash.

**v**) Impregnate with 12mM silver nitrate. gels can turn yellow at this stage, but this does not affect the final quality.

pause point: silver nitrate impregnation can last from 20 minutes to 2 hours without any real change in the quality of the results.

**vi**) Only one gel per tray must be used at this stage. The best setup is to arrange on the bench the box containing the gel(s) soaking in silver nitrate, one box half-filled with water, as many boxes containing developer as gels present in the same box of silver solution, plus one box containing the stop solution. The stop solution, made the day of use, contains 40g of tris and 20ml of acetic acid per liter. With gloved hands, rinsed with deionized water, pull out one gel from the silver solution. Dip it for 10 seconds in the water bath, the pull it out from water and transfer it in the developer solution. A brown or grey precipitate normally develops within a few seconds when the gel is dipped in the developer. This precipitate must be redissolved by shaking of the developer-gel- containing box, other wise a particulate surface background will deposit on the gel surface. When this precipitate has redissolved, repeat the procedure for the next gel of the batch. With this protocol, the most intense bands or spots take a few minutes to appear (up to 5 minutes depending on the temperature). Development can be let to proceed up to background development, but nothing happens after 45 minutes of development.

**vii**) When the adequate degree of staining has been achieved, transfer the gel to the Tris stop solution for at least 30 minutes. Up to four gels can be piled in one box of stop solution.
pause point: time in stop solution can be extended to up to 2 hours

**viii**) wash gels in water (at least, twice, for 30 minutes for each wash). Stained gels can be stored in water for several days, but when mass spectrometry analysis is forecast, much better results are obtained if the staining, washing, spot excision and destaining are performed on the same day [8].

**C) Aldehyde-free silver ammonia staining** [8,13,14] (Timing ca. 5 hours).

**i** Fix the gels in 30% ethanol, 10% acetic acid, and 0.05% naphtalene disulfonic acid. Fixation can be carried out for 3x30 minutes, but also for 30 minutes + overnight, depending on the time frame and of the presence of interfering compounds in the gel (see critical point below)

Pause point: fixation can be carried out for up to 24 hours, with at least one solution change which can take place when desired.

Critical point: Fixation for at least 18 hours is required to remove completely carrier ampholytes used in 2D gels and thus secure a clear background in the low molecular weight region of the gel.

**ii** Rinse the gels in water for 6 x 10 minutes

**iii** Impregnate for 30-60 minutes in silver-ammonia solution.

**iv** Rinse 3 x 5 minutes in water

**v** Develop image (5-10 min) in acidic developer. The most intense bands or spots should appear within 1-2 minutes

**vi** Stop development in EA stop solution. Leave in this solution for 30--60 min.

**vii** Rinse with water (several changes) prior to drying or densitometry.

**D) High sensitivity silver ammonia staining with formaldehyde fixation** [7,13,14] (Timing 2 hours + overnight + 4 hours).

**i** Immediately after electrophoresis, place the gels in water, and let rinse for 5-10 min.

**ii** Soak gels in 20% ethanol containing 10% (v/v) formalin for 1 h.

**iii** Rinse 2x 15 min in water.

**iv** Sensitize overnight in 0.05% naphtalene disulfonate

**v** Rinse 6x 20 min in water.

**vi** Impregnate for 30-60 minutes in silver-ammonia solution.

**vii** Rinse 3 x 5 minutes in water

**viii** Develop image (5-10 min) in acidic developer. The most intense bands or spots should appear within 1-2 minutes

**ix** Stop development in EA stop solution. Leave in this solution for 30--60 min.

**x** Rinse with water (several changes) prior to drying or densitometry.

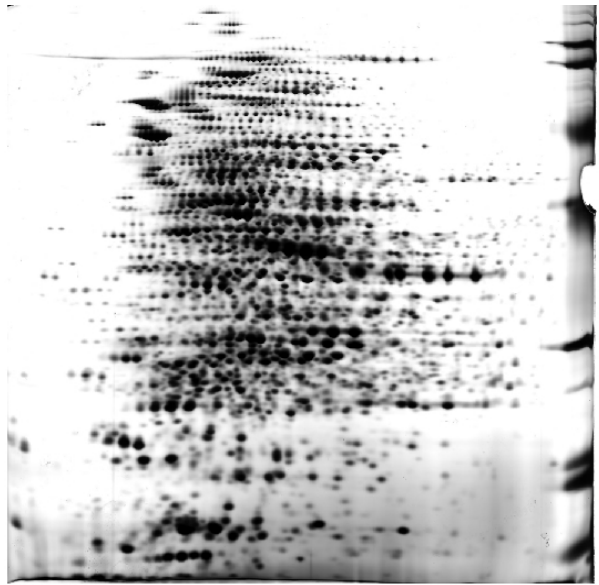 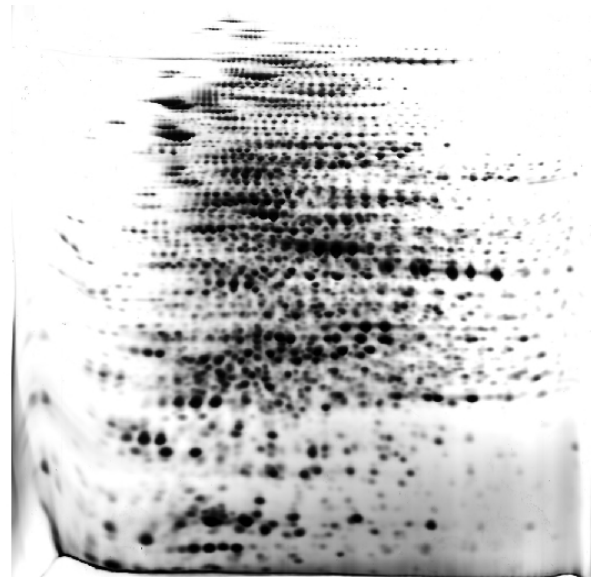

A B

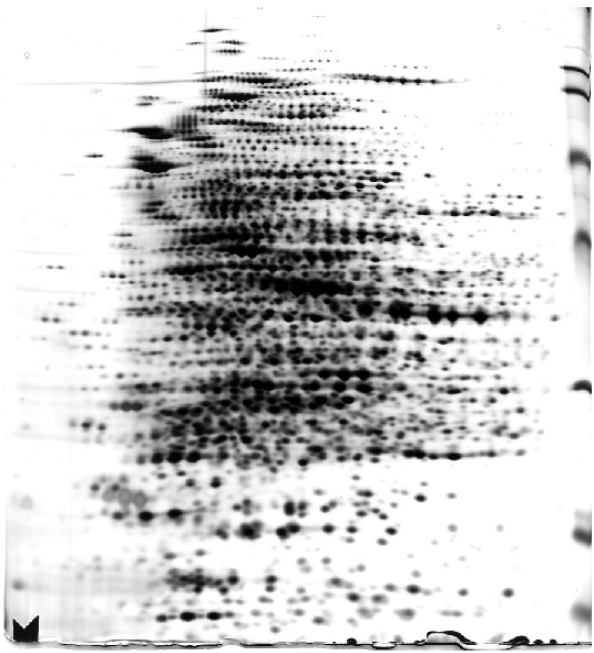 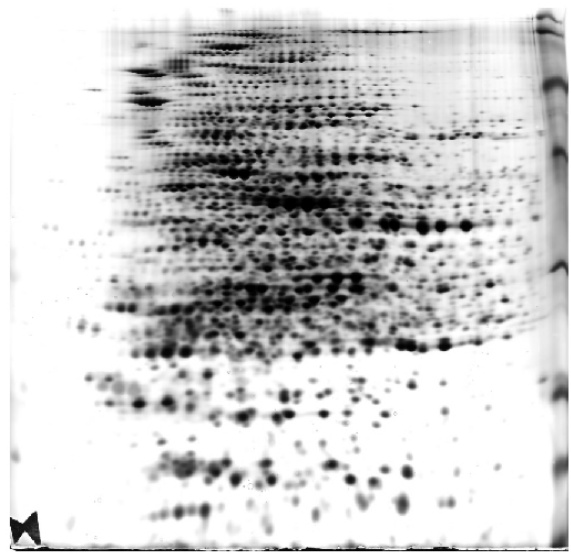

C D

Figure 2: Comparison of the four staining options.
E. coli total proteins (90µg) were separated by 2D electrophoresis (pH 4-8 gradient in the first dimension10% acrylamide for the second dimension.
Detection was carried out by the four staining options described in the text.
A) fast silver nitrate staining; B) long silver nitrate staining; C) aldehyde-free silver-ammonia staining; D) silver-ammonia staining with formaldehyde fixation

**Ancillary protocol: spot destaining prior to mass spectrometry**.
Silver staining interferes strongly with mass spectrometry analysis of spots or bands excised from stained electrophoresis gels [11]. This interference can be reduced by destaining the spots or bands prior to the standard digestion protocols. The destaining protocols giving minimal artifacts is the ferricyanide-thiosulfate protocol of Gharahdaghi et al. [15]. This protocol can be carried out on spots or bands in microtubes (0.5 or 1.5 ml) or in 96 well plates. The use of a shaking device (plate shaker or rotating wheel for tubes) is recommended

Procedure

**i** Cover the spots or bands with 0.15 ml of spot destaining solution. The stain should be removed in 5-10 minutes.

**ii** Remove the solution, and rinse the spots 5x5 minutes with water (0.15 ml per gel piece)

**iii**Remove the water, and soak the gel pieces in 200mM ammonium hydrogenocarbonate (in water) for 20 minutes (0.15 ml per gels piece)

**iv** Repeat step ii above.

Process the rinsed gel pieces, or store dry at –20°C until use

**Troubleshooting**

The reliability and robustness of silver staining has dramatically improved since the early days [2], mostly with the use of thiosulfate as a background reducer [6,8]. However, some problems occur from time to time. Typical examples are given below.

Metallic silver deposits on the surface of the gel (silver mirror).
This is most often due to impurities coming from ill-cleaned glass plates used for gel casting. Better cleaning required. Otherwise, this can be due to pressure marks (finger-like prints). Use only powder-free gloves, and reduce manipulation of the gels to the minimum.

In silver-ammonia staining, the solution cannot be prepared because it makes a brown to black precipitate.
Exhausted ammonia solution. Replace ammonia solution

development does not take place (no or weak image)
there is a mistake either in the developer or in the silver solution. Most common mistakes are wrong silver dilution and use of bicarbonate instead of carbonate. If this occurs on a very precious gel, stop the reaction, rinse with several batches of water (typically 10x30 minutes) and restart from the beginning of the procedure

development is very slow, with weak resulting mage
in addition to the causes mentioned in 3) this can also be due to a too low temperature. Restart as mentioned in 3, but warm water to 18-25°C

in silver-ammonia staining, the gel turns brown in the silvering bath
improper electrophoresis system. Use only the glycine or taurine systems

a strong yellow background appears at the development stage
Exhausted thiosulfate. Change the thiosullfate powder in silver nitrate staining, the developer turns black and makes a deposit on the gel.
a "black cloud" should appear when the gel is dipped in the developer. It should redissolve with shaking. If not, the thiosulfate is exhausted. Change it.

Last but not least, how to save a precious gel with a terrible background
Stop development, rinse. Then destain the gel in a solution containing 3g/l potassium ferricyanide, 6g/l sodium thiosulfate (or even better ammonium thiosulfate) and 10 ml/l concentrated ammonia. This solution must be made just before use and should be green-yellow. When the background has completely disappeared, rinse with several batches of water, until the yellow color of the gel has completely disappeared. It is very likely that the bands or spots will also disappear. Rinse with water for an additional 5x 30 minutes, then perform a fast silver nitrate staining (protocol A), starting directly at the sensitization step (step iiii). Expecting a decent result in mass spectrometry after such an ordeal is somewhat adventurous.

**Anticipated results**

All the silver staining protocols mentioned here allow protein detection in the very low nanogram range (even sub-nanogram), i.e. 30-100 times more sensitive than colloidal Coomassie Blue [16], with very common laboratory equipment and chemicals. The various protocols provide flexibility to meet many different uses, from the fast staining of mini gels to large scale staining of large series of 2D gels. However, this also means that there is no silver staining protocol combining sensitivity, homogeneity of detection and superior compatibility with mass spectrometry.

References

[1] Rabilloud, T. Mechanisms of protein silver staining in polyacrylamide gels: a 10-year synthesis.*Electrophoresis* **11**, 785-794 (1990)

[2] Merril C.R. *et al*. Ultrasensitive stain for proteins in polyacrylamide gels shows regional variation in cerebrospinal fluid proteins. *Science* **211**, 1437-1438 (1981)

[3] Schagger, H. & von Jagow, G. Tricine-sodium dodecyl sulfate-polyacrylamide gel electrophoresis for the separation of proteins in the range from 1 to 100 kDa. *Anal. Biochem.* **166**, 368-79. (1987)

[4] Wiltfang, J. *et al*. A new multiphasic buffer system for sodium dodecyl sulfate-polyacrylamide gel electrophoresis of proteins and peptides with molecular masses 100,000-1000, and their detection with picomolar sensitivity. *Electrophoresis* **12**, 352-366 (1991).

[5] Tastet, C. *et al.* A versatile electrophoresis system for the analysis of high- and low-molecular weight proteins *Electrophoresis* **24**, 1787-1794 (2003)

[6] Hochstrasser, D.F. & Merril, C.R. 'Catalysts' for polyacrylamide gel polymerization and detection of proteins by silver staining. *Appl. Theor. Electrophoresis* **1**, 35-40 (1988)

[7] Eschenbruch, M. & Bürk, R.R. Experimentally improved reliability of ultrasensitive silver staining of protein in polyacrylamide gels. *Anal. Biochem.* **125**, 96-99 (1982)

[8] Chevallet, M. *et al.* Improved mass spectrometry compatibility is afforded by ammoniacal silver staining. *Proteomics* **6**, 2350-2354 (2006)

[9] Blum, H. *et al.* Improved silver staining of plant proteins, RNA and DNA in polyacrylamide gels. *Electrophoresis* **8**, 93-99 (1987)

[10] Rabilloud, T. A comparison between low background silver diammine and silver nitrate protein stains. *Electrophoresis* **13**, 429-439 (1992)

[11] Richert, S. *et al.* About the mechanism of interference of silver staining with peptide mass spectrometry. *Proteomics.* **4**, 909-916 (2004)

[12] Sinha, P. *et al.* A new silver staining apparatus for MALDI/TOF analysis of proteins after two-dimensional electrophoresis *Proteomics* **1**, 835-840 (2001)

[13] Mold, D.E. *et al.* Silver staining of histones in Triton-acid-urea gels. *Anal. Biochem.* **135**, 44-47 (1983)

[14] Rabilloud, T., *et al.* Two-dimensional electrophoresis of human placental mitochondria and protein ientification by mass spectrometry: Toward a human mitochondrial proteome *Electrophoresis* **19**, 1006-1014 (1998)

[15] Gharahdaghi, F., *et al.* Mass spectrometric identification of proteins from silver-stained polyacrylamide gel: a method for the removal of silver ions to enhance sensitivity *Electrophoresis.* **20**, 601-605. (1999)

[16] Neuhoff,V. *et al.* Improved staining of proteins in polyacrylamide gels including isoelectric focusing gels with clear background at nanogram sensitivity using Coomassie Brilliant Blue G-250 and R-250. *Electrophoresis* **9**, 255-262 (1988)